# Current-Induced Helicity Reversal of a Single Skyrmionic Bubble Chain in a Nanostructured Frustrated Magnet

*Zhipeng Hou, Qiang Zhang, Xichao Zhang, Guizhou Xu, Jing Xia, Bei Ding, Hang Li, Senfu Zhang, Nitin M Batra, Pedro MFJ Costa, Enke Liu, Guangheng Wu, Motohiko Ezawa, Xiaoxi Liu[7], Yan Zhou\*, Xixiang Zhang\*, and Wenhong Wang\**

Dr. Z. P. Hou, B. Ding, H. Li, Dr. E. K. Liu, Prof. G. H. Wu, Prof. W. H. Wang
Beijing National Laboratory for Condensed Matter Physics, Institute of Physics, Chinese Academy of Sciences, Beijing 100190, China
E-mail: wenhong.wang@iphy.ac.cn
Dr. Z. P. Hou
Institute for Advanced Materials, South China Academy of Advanced Optoelectronics, South China Normal University, Guangzhou 510006, China
Dr. Q. Zhang, Dr. S. F. Zhang, N. M. Batra, Prof. P. MFJ. Costa, Prof. X. X. Zhang
Physical Science and Engineering Division, King Abdullah University of Science and Technology, Thuwal 23955-6900, Saudi Arabia
E-mail: xixiang.zhang@kaust.edu.sa
Dr. X. C. Zhang, J. Xia, Prof. Y. Zhou
School of Science and Engineering, The Chinese University of Hong Kong, Shenzhen, Guangdong 518172, China
E-mail: zhouyan@cuhk.edu.cn
Dr. G. Z. Xu
School of Materials Science and Engineering, Nanjing University of Science and Technology, Nanjing 210094, China
Prof. M. Ezawa
Department of Applied Physics, The University of Tokyo, 7-3-1 Hongo, Tokyo 113-8656, Japan
Prof. X. X. Liu
Department of Electrical and Computer Engineering, Shinshu University, 4-17-1 Wakasato, Nagano 380-8553, Japan

**Abstract**

**Helicity indicates the in-plane magnetic-moment swirling direction of a skyrmionic configuration. The ability to reverse the helicity of a skyrmionic bubble via purely electrical means has been predicted in frustrated magnetic systems, however its experimental observation has remained challenging. Here, we experimentally demonstrate the current-driven helicity reversal of the skyrmionic bubble in a nanostructured frustrated $Fe_3Sn_2$ magnet. The critical current density required to trigger the helicity reversal is $10^9$ - $10^{10}$ A/m$^2$, with a corresponding pulse-width varying from 1 μs to 100 ns. Computational simulations reveal that both the pinning effect and dipole-dipole interaction play a crucial role in the helicity-reversal process.**

The electrical manipulation of low-dimensional magnetism is a key challenge to better information technology.[1,2] A variety of spin configurations, including magnetic domain walls[3,4] and vortices,[5,6] have been explored for the electrical control of magnetism. Recently, researchers have increasingly focused their interest on the topologically protected vortex-like spin configurations,[7-22] so called magnetic skyrmions or magnetic skyrmionic bubbles, in view of their intriguing electromagnetic properties endowed by the non-trivial topological nature, such as a pinning-free motion with an ultralow current density[11,22-27] and an electric polarization characteristic in insulators.[28-30] Despite these intriguing magneto-electronic functions, the electrical manipulation of the spin texture of a skyrmionic spin configuration, such as the electricity-induced helicity reversal of a skyrmion or skyrmionic bubble, has not yet been well established in experiments.

The helicity indicates the in-plane magnetic-moment swirling direction (e.g., clockwise or counterclockwise) of a skyrmionic spin configuration. In a chiral magnet, the symmetry breaking generates a strong Dzyaloshinskii-Moriya interaction (DMI) that not only stabilizes the skyrmion, but also imprints the chirality of a crystal lattice into the chirality of magnetic orders. This feature makes the skyrmion helicity closely related to the underlying lattice chirality and therefore difficult to be reversed by a purely electrical stimulus, unless the strength of the DMI is reduced to a small value.[4]

Recent studies have demonstrated that some non-chiral centrosymmetric magnets[19-22] or frustrated magnetic systems[31-41] could host skyrmionic bubbles that are stabilized by the interplay of the external magnetic field, exchange interaction/competing exchange interaction, uniaxial magnetic anisotropy, and dipole-dipole interaction (DDI). Skyrmionic bubbles are topologically equivalent to the DMI-stabilized skyrmions. Consequently, the two classes of spin configurations exhibit similar topological properties. However, unlike a DMI-stabilized skyrmion, the skyrmionic bubble does not have a fixed helicity; instead, its helicity possesses an internal degree of freedom taking a binary helicity number $\eta = \pi/2, -\pi/2$ (where $\eta = \pi/2$ represents a counterclockwise swirling direction and $\eta = -\pi/2$ represents a clockwise swirling direction).[32] More importantly, the skyrmionic bubbles with opposite helicities possess the same energy, which potentially enables helicity switching when an external stimulus, such as a spin-polarized current, is applied.[34]

Here, we report experimental observations of the current-driven helicity reversal of skyrmionic bubbles in a nanostructured frustrated $Fe_3Sn_2$ magnet, by using *in-situ* Lorenz transmission electron microscopy (LTEM). The critical current density for triggering the helicity reversal is $10^9$ - $10^{10}$ A/m$^2$, with a corresponding pulse-width varying from 1 μs to 100 ns. By means of simulations, we demonstrate that both the pinning effect and the dipole-dipole interaction (DDI) play a crucial role in the helicity-reversal processes.

Frustrated $Fe_3Sn_2$ magnet has a centrosymmetric rhombohedral structure with an alternate stacking of the Fe-Sn kagome layers along the *c*-axis, as shown in **Figure 1a**. Our previous investigations have experimentally demonstrated that it could host skyrmionic bubbles at room temperature.[41] However, in bulk $Fe_3Sn_2$, the skyrmionic bubbles coexist with trivial bubbles and metastable skyrmionic bubbles, which severely hinders further manipulation of the skyrmionic bubbles via electrical stimuli.[41] Very recently, by implementing a geometrically confined method,[42-46] we realized a single chain of topologically stable skyrmionic bubbles for a wide range of temperatures from 100 K to 630 K,[45] which makes such a investigation feasible.

First, we fabricated nanotrack devices that allowed us to directly study the current-induced dynamics of skyrmionic bubbles in LTEM, from a $Fe_3Sn_2$ single crystal by using focused ion-beam (FIB). Details of the fabrication process can be found in both **Figure S1** and the Methods section. **Figures 1b-c** show the structure of the device, which was composed of a $Fe_3Sn_2$ nanotrack, Si chip, and tungsten (W)-wires. The nanotrack was designed to be 1 μm in width to exclude the trivial bubbles and metastable skyrmionic bubbles in the bulk $Fe_3Sn_2$.[46] The outer parts were coated with amorphous carbon (black region) and platinum (grey region) to protect the $Fe_3Sn_2$ nanotrack during the fabrication process, and to reduce the interfacial Fresnel fringes in the LTEM image (see **Figure 1d**, left panel). A high-angle annular dark-field scanning transmission electron microscopy (HAADF-STEM) image taken from the out-of-plane direction of the nanotrack is presented in the right panel of **Figure 1d**. We found that the atoms were alternately arranged in a kagome lattice, which suggested that the normal direction was along the *c*-axis. To measure the resistance of the device, a DC current was injected along the in-plane direction of the nanotrack. The device exhibited a linear *I–V* dependence, i.e., an ohmic conduction, and the corresponding value of resistance ($R_{xx}$) was calculated to be as low as 45 Ω (see **Figure 1e**). We further investigated the dependence of the domain evolution process on the magnetic field, in the nanotrack, without injection of current. LTEM images, which were taken with different magnetic fields, showed that a densely arranged single chain of skyrmionic bubbles could be created under an external out-of-plane magnetic field *H* of 160 mT (see **Figures 1f-h**). Notably, the uniform helicity of skyrmionic bubbles shown in **Figure 1h** is not a usual case. In a wider field of view or other samples, the helicity was found to be random (see **Figure S2**, SI).

Hereafter, we injected current pulses along the longitudinal direction of the nanotrack. The current density ($j$) ranged from 0 to $4.2\times10^{10}$ A/m$^2$ (the current of 1 mA corresponds approximately to the current density $5\times10^9$ A/m$^2$), and the pulse width ($\tau$) and frequency ($f$) were fixed at 100 ns and 1 Hz, respectively. Since the resistivity of amorphous carbon and platinum are much higher than that of Fe$_3$Sn$_2$, we only considered the current that passed through the Fe$_3$Sn$_2$ layer (see **Figure S3**, SI). The dynamics of the skyrmionic bubbles was recorded as movies, with an exposure time of 50 ms and a frame rate of 60 frames per second (fps). Numerous movies were analyzed to extract information on the changes of the magnetization configuration (see **Figures 2a-e**). For $j$ = 0 to $2.6\times10^{10}$ A/m$^2$, no significant motion or morphology variation of skyrmionic bubbles was observed (see **Figures 2a-b**). However, when $j$ increased to $3\times10^{10}$ A/m$^2$, we observed a discontinuous, current-driven motion of the skyrmionic bubbles, first moving along the nanotrack with an average velocity $v$ of approximately 0.1 m/s and then stopping, in spite of the current pulses being continuously applied (see **Supplementary Movie 1** and **Figures 2c-d**). The cessation of the skyrmionic bubbles may be due to the fact that they were densely arranged along the longitudinal direction of the nanotrack, which resulted in a strong pinning effect on their motion due to the skyrmion-skyrmion and skyrmion-defect interactions.[47,48] By further increasing $j$, no obvious motion was observed anymore. Surprisingly, when $j$ = $3.4\times10^{10}$ A/m$^2$ was injected, the helicity of skyrmionic bubbles began to reverse between the clockwise and counterclockwise directions (see **Supplementary Movie 2** and **Figure 2e**).

To show the details of the helicity reversal process, five sequential snapshots from **Supplementary Movie 2** are presented in **Figures 3a-e**. These images present the changes of the skyrmionic bubbles after a series of current pulses passed the nanotrack and the time-resolved manner of movie allows us to track the helicity variation of a single skyrmion. The individual skyrmions enclosed by the white dashed circles (see **Figures 3a-e**) show a one-to-one correspondence. Since the incident electron beam of LTEM was deflected by the Lorentz force generated by the local in-plane magnetic induction, the spatial variation of the in-plane magnetization led to a bright or dark contrast in the LTEM images. The LTEM image of the skyrmion in its initial state, i.e., without current injection, shows a clear black (core region) and white (edge region) contrast (see **Figure 3a**). By combining the simulations with the LTEM images,[49] we established that the in-plane magnetization swirled counterclockwise (see **Figure 3f**). After the first current pulse, the contrast of the image was completely reversed, i.e., the inner region became white and the outer region became black. This suggests that the in-plane magnetization swirling switched from counterclockwise to clockwise (see **Figure 3g**), meaning that the helicity of the skyrmionic bubble was reversed under electrical stimuli. We also established that the maximum current density for reversing the helicity was $4.2\times10^{10}$ A/m$^2$ (see **Figure S4**, SI), above which the device would be heated beyond the Curie temperature ($T_c$) of Fe$_3$Sn$_2$ ($T_c$ = 640 K). Importantly, we observed the current-induced helicity reversal in different devices (see **Supplementary Movie 3**), which suggested that this intriguing phenomenon was of solid physical origin. **Figure 3h** summarizes the helicity variation of the enclosed skyrmionic bubbles in **Figure. 3a** in term of the current pulse number $N$. We found that the helicity did not reverse after a certain number of pulses, e.g., after the third current pulse ($N$ = 3), the helicity remained unchanged. The random skipping feature may be due to a local fluctuation of energy or the pinning effect, which was closely related to the inhomogeneity of the sample and the defects possibly introduced during the crystal growth process and/or the FIB fabrication . In addition to helicity reversal, the current pulse also induced a mutual transformation between stripes and skyrmionic bubbles (enclosed by yellow boxes in **Figures 3b-d**). This feature was similar to that observed in the DMI-stabilized skyrmions[27] and may be attributed to the skyrmion-skyrmion interactions that depended on the distance and relative helicity of adjacent skyrmions.[34,50] The current density threshold for reversing the helicity ($j_{th}$) was about $10^{10}$ A/m$^2$ for a pulse width of 100 ns. When we increased the pulse width, $j_{th}$ decreased. For example, $j_{th}$ decreased to $10^9$ A/m$^2$, when $\tau$ = 1 μs (see **Figure S5**, SI). This value was two or three orders of magnitude lower than that required to drive a conventional domain wall or Néel-type skyrmions,[13-15] but several orders of magnitude higher than that required to manipulate skyrmions in chiral magnets.[11,25,27,51] The higher current density is mainly due to the smaller pulse width, higher energy needed for helicity reversal, and larger size of skyrmionic bubbles in our experiments.

To better understand the experimentally observed helicity reversal process, we simulated the current-driven dynamics on basis of the experimental parameters for a Fe$_3$Sn$_2$ single crystal. Simulation details are provided in the **Supplementary Note 1**. First, we simulated the magnetization dynamics in the absence of a current pulse by considering a classical Heisenberg model with exchange frustration (see **Figure S6**, SI). The simulated results were found in good agreement with the experimental observations, which thus validated our theoretical model and numerical approach. Subsequently, we simulated the current-driven dynamics of an individual skyrmionic bubble by considering the adiabatic and non-adiabatic spin transfer torques (STTs) on basis of the Zhang-Li model.[52] Namely, the conduction electrons were directly spin-polarized by the local magnetic moment and followed the local magnetization direction in the adiabatic limit. We found that the helicity reversal was closely related to the pinning effect. For a weaker pinning center (e.g., the skyrmion core is initially pinned by an anisotropy $K_{pinning} = 3K_u$, where $K_u$ is the uniaxial magnetocrystalline anisotropy constant), the skyrmionic bubble de-pinned and moved within the sample when a current pulse was applied (see **Figures 4a-e** and **Supplementary Movie 4**). In such a process, no helicity reversal was observed though the skyrmionic bubble was initially distorted due to the pinning effect. In contrast, when the core was strongly pinned ($K_{pinning} = 7K_u$), the skyrmionic bubble could not move anymore but its helicity started to reverse from $\eta = \pi/2$ to $-\pi/2$, with a certain distortion and fluctuation of the spin texture (see **Figures 4f-j** and **Supplementary**

**Movie 5**). It is important to note that we have also simulated this type of current-driven dynamics by considering a three-dimensional model or a chain of skyrmionic bubbles, and that similar results were obtained (see **Figures S7-9**, SI). These simulations suggested that the motion of a skyrmionic bubble is a better option than the helicity reversal in energy (see **Figure S10**, SI), which is consistent with our experimental observations. Additionally, during such a helicity reversal process, the skyrmion number ($Q$) showed a sharp transition of $1 \rightarrow 0 \rightarrow 1$ (see **Figure 4k**), meaning that the in-plane spin structure of the skyrmionic bubble was first destructed and then re-formed. This process is different from a typical STT-switching process where the spin structure should not be destructed.[53,54] However, it is also distinctly different from a simple destruction and reformation of equilibrium because the helicity of skyrmionic bubble in our simulations tends to be reversed rather than randomly reforms. We may class such a process as a kind of STT-induced or STT-guided helicity switching accompanying a destruction and reformation of skyrmionic bubble. On the other hand, we found that DDI, i.e. the demagnetization, played a crucial role in the helicity reversal process. We simulated the total energy of a static skyrmionic bubble as a function of $\eta$ (see **Figure 4l**), assuming that the skyrmionic bubble had the same spin profile (i.e., the diameter equals 85 nm) but different helicities. When the DDI was excluded, the energy of the skyrmionic bubble was independent of $\eta$, and both the Néel-type and Bloch-type skyrmionic bubbles were energetically identical. However, when the DDI was included, the formation of the Bloch-type skyrmionic bubble was favored, as they had a much lower energy (i.e., global energy minimum). This feature suggested that the DDI played a dominant role in stabilizing the Bloch-type skyrmionic bubbles. Moreover, the DDI made the helicity of skyrmionic bubbles possess an internal degree of freedom (see **Figure 4l**). Namely, the skyrmionic bubbles with the helicities of $\pi/2$ and $-\pi/2$ possess the same energy (see **Figure 4l**). Hence, it is reasonable to think that the helicity may be freely reversed via STT, when the energy injected by the current pulse is large enough to overcome the energy barrier between the two stable states ($\eta = \pm\pi/2$).

Our experimental results showed that the helicity reversal of pinned skyrmionic bubbles could occur in combination with the annihilation and de-pinning motion of skyrmionic bubbles, and skyrmion-stripe conversion (see **Supplementary Movie 2**). The simulations presented in **Figure 4** and **Figures. S8-9** clearly show that, when the skyrmionic bubbles are pinned by the weak ($K_{pinning}= 3K_u$) or strong ($K_{pinning} = 7K_u$) pinning centers, a de-pinning motion or robust helicity reversal could occur in a chain of skyrmionic bubbles. Further simulations were performed by considering pinning sites with mediate values of $K_{pinning}$ ($3K_u< K_{pinning}< 7K_u$) and different distances between two pinning sites (see **Figure 5**). Both the annihilation ($K_{pinning} = 5K_u$) and stochastic helicity switching of skyrmionic bubbles ($K_{pinning} = 6K_u$) could be obtained via tuning the strength of the pinning centers (see **Figures 5a-h**). Moreover, the distance between two neighbor skyrmionic bubbles also appeared to have a significant influence on the variations of the spin configurations. If the distance between two neighbor skyrmionic bubbles was decreased from 250 nm (the distance between two pinning centers in **Figures 5a-h** and **Figures S8-9** was set at 250 nm) to 200 nm, a skyrmion-stripe conversion could be achieved (**Figures 5i-l**). By tuning the strength and distance of pinning centers in a chain of skyrmions, we could mimic all the variations of spin texture observed in the experiments, which strongly suggests that such stochastic variations originate from the random distribution of pinning centers in the nanotrack that was indeed a very reasonable situation in the real materials.

Finally, we would like to discuss the effects of local heating on the helicity reversal. It has been well-established that the injected current pulse could induce a Joule heating. The corresponding thermal effect may also lead to the helicity reversal of skyrmionic bubbles, as observed in BaFeScMgO,[55] where the helicity starts to reverse when the temperature of the sample is increased to approximately $T_c$. Our simulations demonstrated that, when a current pulse of $j = 3.4 \times 10^{10}$ A/m$^2$ and $\tau = 100$ ns was injected into the nanotrack, the highest temperature that the sample could be heated up to was approximately 480 K, obviously far below $T_c$ of Fe$_3$Sn$_2$ (see **Supplementary Note 2** and **Figure S11**). Our previous investigations demonstrated that the skyrmionic bubbles in the Fe$_3$Sn$_2$ nanotrack have an good temperature stability, i.e. their size, morphology, and helicity remained unchanged across a wide range of temperatures from 300 K to 630 K.[37] Hence, we propose that the observed helicity reversal does not directly originate from the thermal effect but STT. Although the thermal effect is not the dominant factor for the helicity reversal in our samples, it is beneficial for the reversal process, as the thermal energy could lower the effective energy barrier to be overcome in the reversal process.[56] This hypothesis could be validated by our experimental results that showed a decrease of the critical current density as the current pulse-width increased.

In this work, we investigated the current-induced dynamics of skyrmionic bubbles in the nanostructured frustrated magnet Fe$_3$Sn$_2$, using both LTEM observations and computational simulations. We found that the helicity of the skyrmionic bubbles could be electrically reversed by a spin-polarized current along the in-plane direction of the nanotrack. The corresponding critical current density was about $10^9$ - $10^{10}$ A/m$^2$ with a pulse-width ranging from 100 ns to 1μs. Computational simulations revealed that both the local pinning effect and DDI played crucial roles in the helicity reversal. Our results offered valuable insights into the fundamental mechanisms underlying the current-induced dynamics of skyrmionic bubbles.

**Experimental section**

*Sample preparation:* The Fe$_3$Sn$_2$ nanotrack device for *in-situ* LTEM characterization was fabricated by a FIB-SEM dual-beam system. (i) A lamella (thickness of ~3 μm) was caved on the (001) surface of a Fe$_3$Sn$_2$ single crystal by FIB milling. After further fine thinning, the final lamella was 1 μm in thickness. (ii) Layers of C and Pt were deposited on both side of the Fe$_3$Sn$_2$ lamella by using a GIS system to protect the edge of the nanostripe during the nano-manipulation process. (iii) A cuboid was cut from the lamella by FIB milling and lifted out with an Omniprobe nanomanipulator. (iv) The cuboid was transferred onto a customized silicon chip and attached to the electrodes by tungsten deposition using the GIS system. The electrodes of the silicon chip were parallel to the horizontal plane. (v) The silicon chip was rotated 90 degree (the electrodes of the silicon chip were perpendicular to the horizontal plane). Then, the cuboid was thinned to 200 nm along the vertical plane.

*LTEM measurements:* The magnetic domain structure was detected using a Titan G2 60–300 (FEI), in the Lorentz TEM mode, equipped with a spherical aberration corrector for an imaging system, at an acceleration voltage of 300 kV. The objective lens was turned off when the sample holder was inserted, and the perpendicular magnetic field was applied to the sample by increasing the objective lens, gradually, in very small increments.

**Supporting Information**

Supporting Information is available from the Wiley Online Library or from the author.

**Acknowledgements**

This work was supported by the National Key R&D Program of China (Grant No. 2017YFA0303202), the National Natural Science Foundation of China (Grant Nos. 11574137, 11604148, 11874410, 11974298 and 61961136006), the King Abdullah University of Science and Technology (KAUST) Office of Sponsored Research (OSR) under Award No. CRF-2015-2549-CRG4 and No. 2016-CRG5-2977, the Presidential Postdoctoral Fellowship and President's Fund of CUHKSZ, Longgang Key Laboratory of Applied Spintronics, the Shenzhen Fundamental Research Fund (Grant No. JCYJ20170410171958839), Shenzhen Peacock Group Plan (Grant No. KQTD20180413181702403), the Key Research Program of the Chinese Academy of Sciences (Grant No. KJZD-SW-M01), the Grants-in-Aid for Scientific Research from JSPS KAKENHI (Grant Nos. JP18H03676, JP17K05490,

JP15H05854 and JP17K19074), and CREST, JST (Grant Nos. JPMJCR1874 and JPMJCR16F1). Z. P. Hou, Q. Zhang, and X. Zhang contributed equally to this work.

**References**

[1] C. Chappert, A. Fert, F. N. Van Dau, *Nat. Mater.* **2007**, *6*, 813.

[2] J. A. Katine, E. E. Fullerton, *J. Magn. Magn. Mater.* **2008**, *320*, 1217.

[3] S. S. P. Parkin, M. Hayashi, L. Thomas, *Science* **2008**, *320*, 190.

[4] G. V. Karnad, F. Freimuth, E. Martinez, R. Lo Conte, G. Gubbiotti, T. Schulz, S. Senz, B. Ocker, Y. Mokrousov, M. Kläui, *Phys. Rev. Lett.* **2018**, *121*, 147203.

[5] Q. Mistral, M. van Kampen, G. Hrkac, Joo-Von Kim, T. Devolder, P. Crozat, C. Chappert, L. Lagae, T. Schrefl, *Phys. Rev. Lett.* **2008**, *100*, 257201.

[6] A. Yamaguchi, T. Ono, S. Nasu, K. Miyake, K. Mibu, T. Shinjo, *Phys. Rev. Lett.* **2004**, *92*, 077205.

[7] A. Fert, V. Cros, J. Sampaio, *Nat. Nanotech.* **2013**, *8*, 152.

[8] A. Rosch, *Nat. Nanotech.* **2013**, *8*, 160.

[9] R. Wiesendanger, *Nat. Rev. Mater.* **2016**, *1*, 16044.

[10] W. J. Jiang, P. Upadhyaya, W. Zhang, G. Q. Yu, M. B. Jungfleisch, F. Y. Fradin, J. E. Pearson, Y. Tserkovnyak, K. L. Wang, O. Heinonen, S. G. E. te Velthuis, A. Hoffmann, *Science* **2015**, *349*, 283.

[11] D. Liang, J. P. DeGrave, M. J. Stolt, Y. Tokura, S. Jin, *Nat. Commun.* **2015**, *6*, 8217.

[12] S. Woo, K. Litzius, B. Krüger, M. -Y. Im, L. Caretta, K. Richter, M. Mann, A. Krone, R. M. Reeve, M. Weigand, P. Agrawal, I. Lemesh, M. -A. Mawass, P. Fischer, M. Kläui, G. S. D. Beach, *Nat. Mater.* **2016**, *15*, 501.

[13] L. Caretta, M. Mann, F. Büttner, K. Ueda, B. Pfau, C. M. Günther, P. Hessing, A. Churikova, C. Klose, M. Schneider, D. Engel, C. Marcus, D. Bono, K. Bagschik, S. Eisebitt, G. S. D. Beach, *Nat. Nanotech.* **2018**, *13*, 1154.

[14] F. Büttner, I. Lemesh, M. Schneider, B. Pfau, C. Günther, P. Hessing, J. Geilhufe, L. Caretta, D. Engel, B. Krüger, J. Viefhaus, S. Eisebitt, G. S. D. Beach, *Nat. Nanotech.* **2017**, *12*, 1040.

[15] S. Woo, K. M. Song, X. C. Zhang, Y. Zhou, M. Ezawa, X. X. Liu, S. Finizio, J. Raabe, N. J. Lee, S. Kim, S. -Y. Park, Y. Kim, J. -Y. Kim, D. Lee, O. Lee, J. W. Choi, B. Min, H. C. Koo, J. Chang, *Nat. Commun.* **2018**, *9*, 959.

[16] D. Maccariello, W. Legrand, N. Reyren, K. Garcia, K. Bouzehouane, S. Collin, V. Cros, A. Fert, *Nat. Nanotech.* **2018**, *13*, 233.

[17] A. Hrabec, J. Sampaio, M. Belmeguenai, I. Gross, R. Weil, S. M. Chérif, A. Stashkevich, V. Jacques, A. Thiaville, S. Rohart, *Nat. Commun.* **2017**, *8*, 15765.

[18] G. Q. Yu, P. Upadhyaya, X. Li, W. Y. Li, S. K. Kim, Y. B. Fan, K. L. Wong, Y. Tserkovnyak, P. K. Amiri, K. L. Wang, *Nano Lett*. **2016**, *16*, 1981.

[19] W. H. Wang, Y. Zhang, G. Z. Xu, L. C. Peng, B. Ding, Y. Wang, Z. P. Hou, X. M. Zhang, X. Y. Li, E. K. Liu, S. G. Wang, J. W. Cai, F. W. Wang, J. Q. Li, F. X. Hu, G. H. Wu, B. G. Shen, X. -X. Zhang, *Adv. Mater.* **2016**, *28*, 6887.

[20] X. Z. Yu, M. Mostovoy, Y. Tokunaga, W. Z. Zhang, K. Kimoto, Y. Matsui, Y. Kaneko, N. Nagaosa, Y. Tokura, *Proc. Natl. Acad. Sci.USA* **2012**, *109*, 8856.

[21] X. Z. Yu, Y. Tokunaga, Y. Taguchi, Y. Tokura, *Adv. Mater.* **2017**, *29*, 1603958.

[22] X. Z. Yu, Y. Tokunaga, Y. Kaneko, W. Z. Zhang, K. Kimoto, Y. Matsui, Y. Taguchi, Y. Tokura, *Nat. Commun.* **2014**, *5*, 3198.

[23] J. Iwasaki, M. Mochizuki, N. Nagaosa, *Nat. Commun.* **2013**, *4*, 1463.

[24] J. Iwasaki, M. Mochizuki, and N. Nagaosa, *Nat. Nanotech.* **2013**, *8*, 742.

[25] F. Jonietz, S. Mühlbauer, C. Pfleiderer, A. Neubauer, W. Münzer, A. Bauer, T. Adams, R. Georgii, P. Böni, R. A. Duine, K. Everschor, M. Garst, A. Rosch, *Science* **2010**, *330*, 1648.

[26] J. D. Zang, M. Mostovoy, J. H. Han, N. Nagaosa, *Phys. Rev. Lett.* **2011**, *107*, 136804.

[27] X. Z. Yu, D. Morikawa, Y. Tokunaga, M. Kubota, T. Kurumaji, H. Oike, M. Nakamura, F. Kagawa, Y. Taguchi, T. Arima, M. Kawasaki, Y. Tokura, *Adv. Mater.* **2017**, *29*, 1606178.

[28] S. Seki, X. Z. Yu, S. Ishiwata, Y. Tokura, *Science* **2012**, *336*, 198.

[29] J. S. White, K. Prša, P. Huang, A. A. Omrani, I. Živković, M. Bartkowiak, H. Berger, A. Magrez, J. L. Gavilano, G. Nagy, J. Zang, H. M. Rønnow, *Phys. Rev. Lett*. **2014**, *113*, 107203.

[30] P. Huang, M. Cantoni, A. Kruchkov, J. Rajeswari, A. Magrez, F. Carbone, H. M. Rønnow, *Nano Lett.* **2018**, *18*, 5167.

[31] T. Okubo, S. Chung, H. Kawamura, *Phys. Rev. Lett.* **2012**, *108*, 017206.

[32] A. O. Leonov, M. Mostovoy, *Nat. Commun.* **2015**, *6*, 8275.

[33] A. Q. Leonov, M. Mostovoy, *Nat. Commun.* **2017**, *8*, 14394.

[34] X. C. Zhang, J. Xia, Y. Zhou, X. X. Liu, H. Zhang, M. Ezawa, *Nat. Commun*. **2017**, *8*, 1717.

[35] S. Hayami, S. -Z. Lin, C. D. Batista, *Phys. Rev. B* **2016**, *93*, 184413.

[36] S.-Z. Lin, S. Hayami, C. D. Batista, *Phys. Rev. Lett.* **2016**, *116*, 187202.

[37] C. D. Batista, S.-Z. Lin, S. Hayami, Y. Kamiya, *Rep. Prog. Phys.* **2016**, *79*, 84504.

[38] H. Y. Yuan, O. Gomonay, M. Kläui, *Phys. Rev. B* **2017**, *96*, 134415.

[39] L. Rózsa, A. Deák, E. Simon, R. Yanes, L. Udvardi, L. Szunyogh, U. Nowak, *Phys. Rev. Lett.* **2016**, *117*, 157205.

[40] P. Sutcliffe, *Phys. Rev. Lett*. **2017**, *118*, 247203.

[41] Z. P. Hou, W. J. Ren, B. Ding, G. Z. Xu, Y. Wang, B. Yang, Q. Zhang, Y. Zhang, E. K. Liu, F. Xu, W. H. Wang, G. H. Wu, X. X. Zhang, B. G. Shen, Z. D. Zhang, *Adv. Mater.* **2017**, *29*, 1701144.

[42] H. F. Du, R. C. Che, L.Y. Kong, X. B. Zhao, C. M. Jin, C. Wang, J. Y. Yang, W. Ning, R. W. Li, C. Q. Jin, X. H. Chen, J. D. Zang, Y. H. Zhang, M. L. Tian, *Nat. Commun.* **2018**, *6*, 8504.

[43] M. J. Stolt, Z. -A. Li, B. Phillips, D. S. Song, N. Mathur, R. E. Dunin-Borkowski, S. Jin, *Nano Lett.* **2017**, *17*, 508.

[44] H. F. Du, J. P. DeGrave, F. Xue, D. Liang, W. Ning, J. Y. Yang, M. L. Tian, Y. H. Zhang, S. Jin, *Nano Lett.* **2014**, *14*, 2026.

[45] Z. P. Hou, Q. Zhang, G. Z. Xu, C. Gong, B. Ding, Y. Wang, H. Li, E. K. Liu, F. Xu, H. W. Zhang, Y. Yao, G. H. Wu, X. -X. Zhang, W. H. Wang, *Nano Lett.* **2018**, *18*, 1274.

[46] Z. P. Hou, Q. Zhang, G. Z. Xu, S. F. Zhang, C. Gong, B. Ding, H. Li, F. Xu, Y. Yao, E. K. Liu, G. H. Wu, X. -X. Zhang, W. H. Wang, *ACS Nano* **2019**, *13*, 922.

[47] X. C. Zhang, G. P. Zhao, H. Fangohr, J. P. Liu, W. X. Xia, J. Xia, F. J. Morvan, *Sci. Rep.* **2015**, *5*, 7643.

[48] C. Reichhardt, C. J. O. Reichhardt, *Rep. Prog. Phys.* **2017**, *80*, 026501.

[49] J. Cui, Y. Yao, X. Shen, Y. G. Wang, R. C. Yu, *J. Magn. Magn. Mater.* **2018**, *454*, 304.

[50] H. F. Du, X. B. Zhao, F. N. Rybakov, A. B. Borisov, S. S. Wang, J. Tang, C. M. Jin, C. Wang, W. S. Wei, N. S. Kiselev, Y. H. Zhang, R. C. Che, S. Blügel, M. L. Tian, *Phys. Rev. Lett.* **2018**, *120*, 197203.

[51] K. Shibata, T. Tanigaki, T. Akashi, H. Shinada, K. Harada, K. Niitsu, D. Shindo, N. Kanazawa, Y. Tokura, T. Arima, *Nano Lett.* **2018**, *18*, 929.

[52] S. Zhang, Z. Li, *Phys. Rev. Lett.* **2004**, *93*, 127204.

[53] R. H. Koch, J. A. Katine, J. Z. Sun, *Phys. Rev. Lett.* **2004**, *92*, 088302.

[54] J. Z. Sun, T. S. Kuan, J. A. Katine, R. H. Koch, *Proc. SPIE* **2004**, *5359*, 445.

[55] Y. M. Huai, *AAPPS Bull.* **2008**, *18*, 33.

[56] X. Z. Yu, K. Shibata, W. Koshibae, Y. Tokunaga, Y. Kaneko, T. Nagai, K. Kimoto, Y. Taguchi, N. Nagaosa, Y. Tokura, *Phys. Rev. B* **2016**, *93*, 134417.

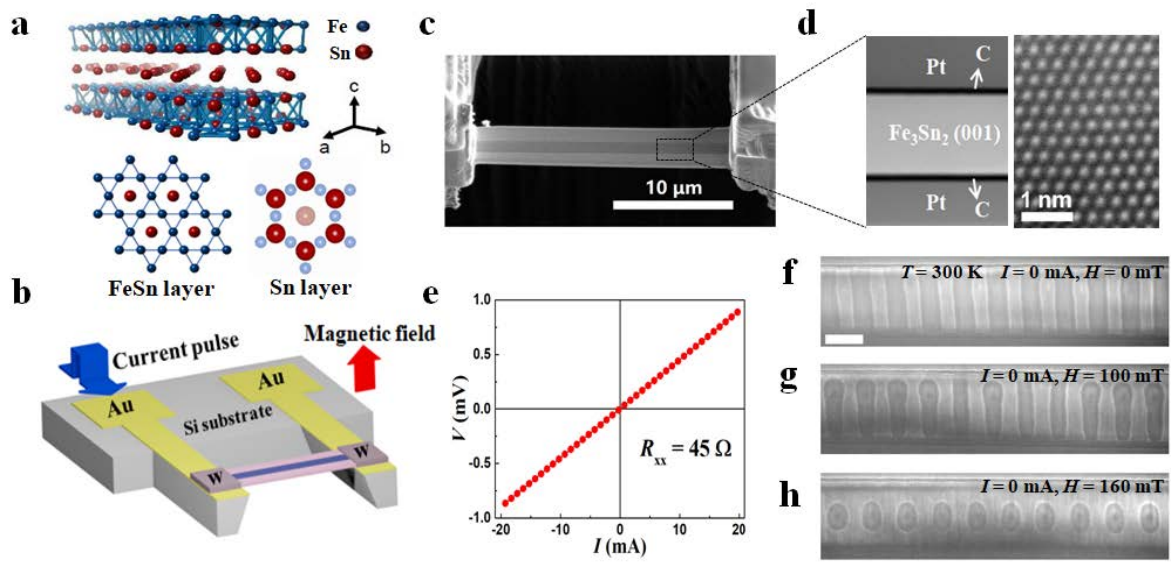

**Figure 1.** Structure and domain evolution of nanotrack device. **a** Crystal structure of $Fe_3Sn_2$ (upper); top views of the kagome lattice of FeSn layer (lower, left) and the Sn layer (lower, right). **b** Schematic of the device. **c** Scanning electron microscopy (SEM) image of the device. **d** The left panel shows a scanning transmission electron microscopy (STEM) image of the region from **c**. The right panel shows the HAADF-STEM image. **e** $I$–$V$ curve for the device. **f-h** Magnetic field dependence of LTEM images with a current density of 0 A/m$^2$. The scale bar is 500 nm.

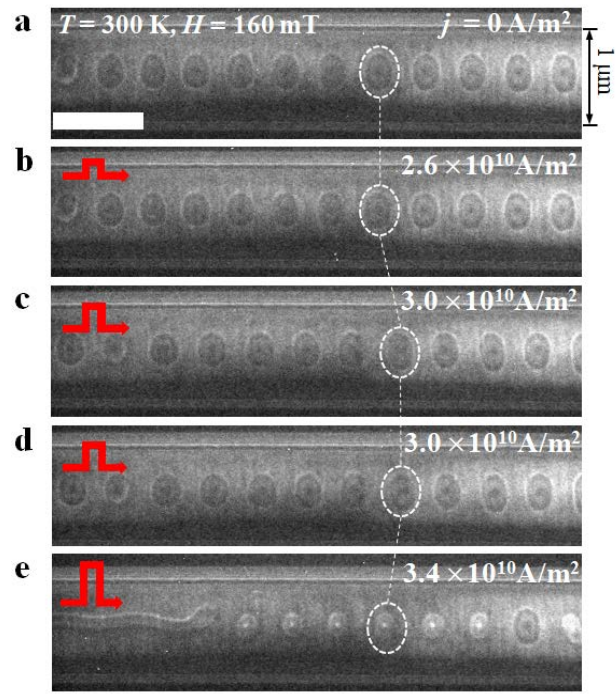

**Figure 2.** Current-driven motion and helicity reversal of skyrmionic bubbles. **a** LTEM image without injection of current pulse. LTEM images after injecting current pulse at a density of **b** $2.6\times10^{10}$ A/m$^2$, **c-d** $3.0\times10^{10}$ A/m$^2$, and **e** $3.4\times10^{10}$ A/m$^2$. The current pulses possess a fixed pulse width of 100 ns and frequency of 1 HZ. The scale bar is in **a** 1 μm.

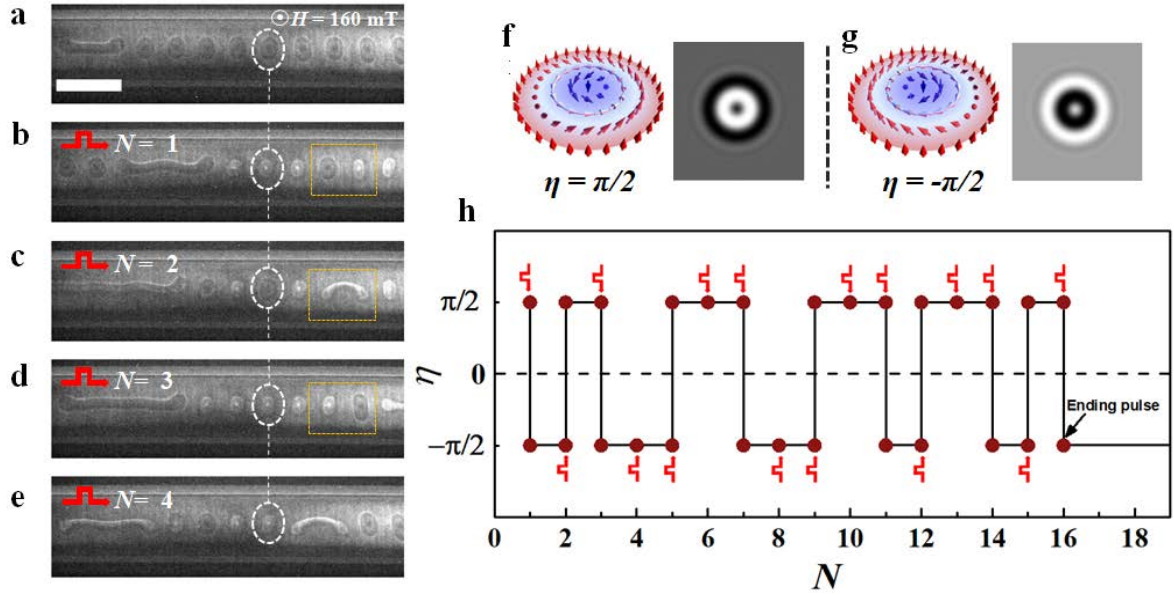

**Figure 3.** Current-driven helicity reversal of skyrmionic bubbles. **a-e** Sequential LTEM images with $H$ = 160 mT after injecting current pulses with $j = 3.4 \times 10^{10}$ A/m$^2$, $\tau$ = 100 ns and $f$ = 1 Hz. The skyrmions enclosed by white circles show a one-to-one correspondence. Domains enclosed by the yellow boxes in **b-d** show the stripe-skyrmion conversion. The red arrows in **b-e** represent the direction of the injected current flow. The left panel of **f** and **g** show the schematic view of two skyrmions with opposite helicity, respectively. The right panels of **f** and **g** are their corresponding simulated LTEM images based on their spin texture. **h** Helicity reversal of the skyrmionic bubble enclosed by white circles with respect to the current pulse number $N$. The scale bar is in **a** 1 μm.

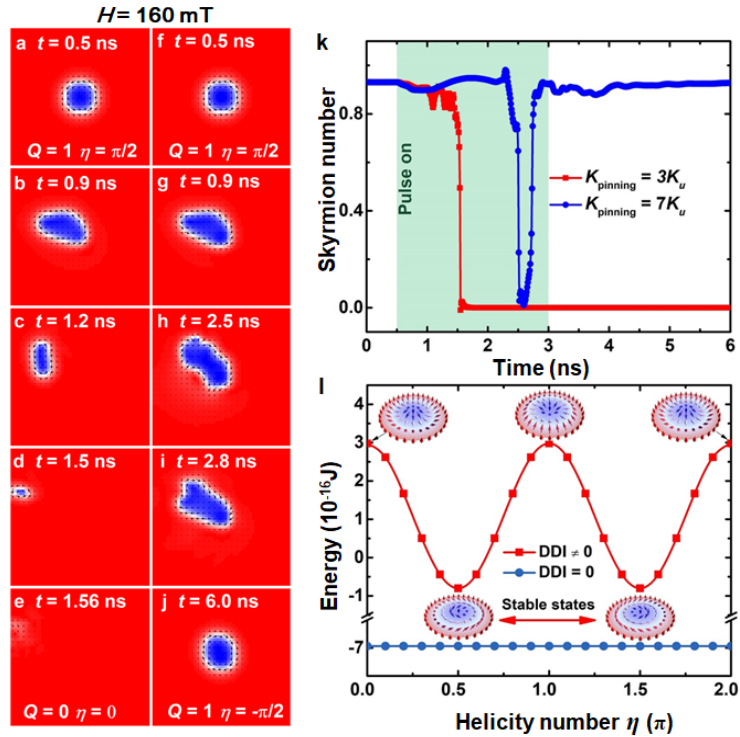

**Figure 4.** Simulation of current-induced helicity reversal. **a-e** Snapshots at five selected times, where the skyrmion core is pinned by an anisotropy $K_{pinning} = 3K_u$. A spin-polarized current of $j = -1.5 \times 10^{12}$ A/m$^2$ is injected during $t = 0.5 \sim 3$ ns. **f-j** Snapshots at five selected times showing the helicity reversal of a skyrmionic bubble driven by the same current pulse as that in **a**, where the skyrmion core is initially pinned due to a higher anisotropy $K_{pinning} = 7K_u$. **k** Skyrmion number as a function of time. **l** The total micromagnetic energy of a skyrmion as a function of $\eta$ with $H = 160$ mT.

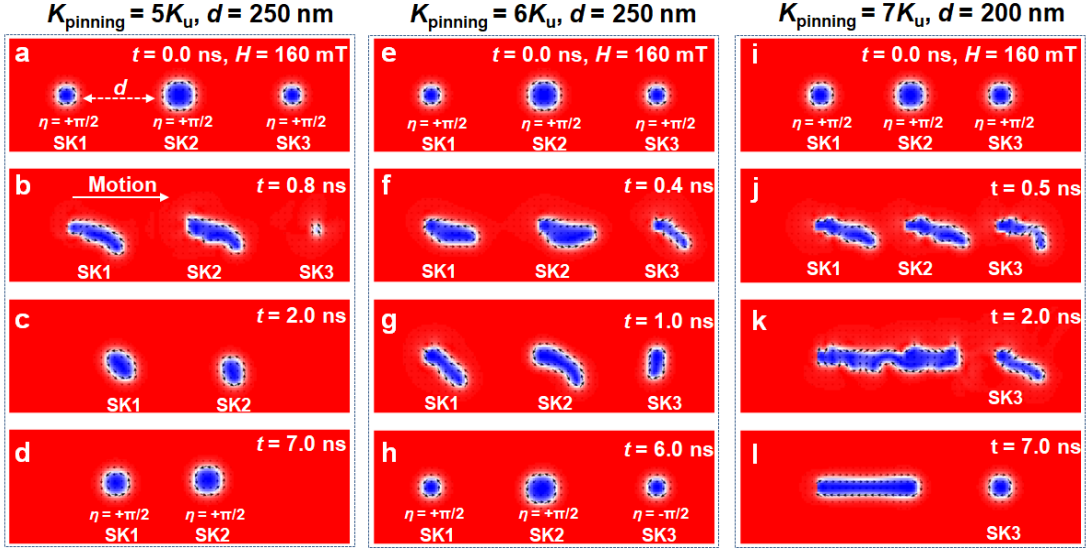

**Figure 5.** Simulations of current-induced variation of spin textures with different strength $K_{\text{pinning}}$ and distances $d$ of pinning centers. **a-d** Snapshots at selected times, where the skyrmion cores are pinned by an anisotropy $K_{\text{pinning}} = 5K_u$ with $d$ of 250 nm. The spin current of $j = 2.2\times10^{12}$ A m$^{-2}$ is injected. The total simulation time is 7.0 ns. After the injection of current pulse, Sk3 is destroyed, while Sk1 and Sk2 are unpinned and driven into motion as the pinning sites are not strong enough to prevent their motion. The arrow indicates motion direction. **e-h** Snapshots at selected times showing the helicity reversal of a skyrmionic bubble chain driven by a spin current of $j = 2.2\times10^{12}$ A m$^{-2}$. The skyrmion cores are initially pinned by $K_{\text{pinning}} = 6K_u$ with $d$ of 250 nm. The total simulation time is 6.0 ns. After the injection of current pulse, the helicities of Sk3 is reversed from $\eta = +\pi/2$ to $\eta = -\pi/2$, while the helicity of Sk1 and Sk2 remain unchanged as $\eta = +\pi/2$. **i-l** Snapshots at selected times showing the helicity reversal of a skyrmionic bubble chain driven by a spin current of $j = 2.2\times10^{12}$ A m$^{-2}$. The skyrmion cores are initially pinned by $K_{\text{pinning}} = 7K_u$ with $d$ of 200 nm. The total simulation time is 7.0 ns. After the injection of current pulse, Sk1 and Sk2 form a stripe domain, while the helicity of Sk3 is reversed from $\eta = +\pi/2$ to $\eta = -\pi/2$. The external magnetic field in **a-l** is fixed to be 160 mT.

**The current-driven helicity revesal of skyrmionic bubbles** is realized for the first time in a nanostructured frustrated magnet $Fe_3Sn_2$. The critical current density required to trigger the helicity reversal is about $10^9$ - $10^{10}$ A/m$^2$. Our results offer valuable insights into the fundamental mechanisms underlying the current-induced dynamics of skyrmionic bubbles.

**Keywords:** skyrmions, frustrated magnet, current-induced helicity reversal, spin-polarized current

*Zhipeng Hou, Qiang Zhang, Xichao Zhang, Guizhou Xu, Jing Xia, Bei Ding, Hang Li, Senfu Zhang, Nitin M Batra, Pedro MFJ Costa, Enke Liu, Guangheng Wu, Motohiko Ezawa, Xiaoxi Li[7], Yan Zhou\*, Xixiang Zhang\*, and Wenhong Wang\**

**Current-Induced Helicity Reversal of a Single Skyrmionic Bubble Chain in a Nanostructured Frustrated Magnet**

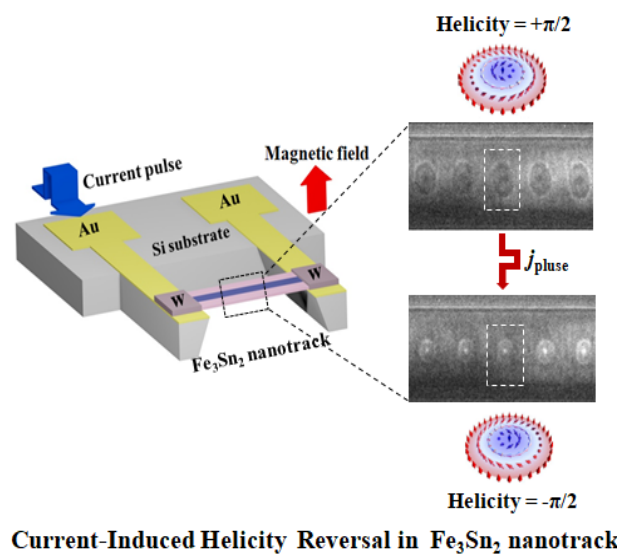

Table of Content

21